\documentclass[prd,showpacs,floats,floatfix,nofootinbib]{revtex4}
\usepackage{color}
\usepackage{amsbsy}
\usepackage{amssymb}
\usepackage{amsmath}
\input{epsf}
\usepackage{graphicx}

\usepackage{amssymb}
\usepackage{bm}
\def\spose#1{\hbox to 0pt{#1\hss}}

\def\lta{\mathrel{\spose{\lower 3pt\hbox{$\mathchar"218$}}
     \raise 2.0pt\hbox{$\mathchar"13C$}}}
\def\gta{\mathrel{\spose{\lower 3pt\hbox{$\mathchar"218$}}
     \raise 2.0pt\hbox{$\mathchar"13E$}}}

\def\beq{\begin{equation}}
\def\eeq{\end{equation}}
\def\bea{\begin{eqnarray}}
\def\eea{\end{eqnarray}}

\def\s{{\rm s}}

\begin{document}
\title{Using gravitational-wave data to constrain dynamical tides in neutron star binaries}

\author{N. Andersson$^1$ \& W.C.G. Ho$^{1,2}$}
\affiliation{$^1$ Mathematical Sciences and STAG Research Centre, University of Southampton, Southampton \\ $^2$ School of Physics and Astronomy, University of Southampton, Southampton 
SO17 1BJ, UK}

\date{\today}

\begin{abstract}
We discuss the role of dynamical tidal effects for inspiralling neutron star binaries, focussing on features that may be considered ``unmodelled'' in gravitational-wave searches. In order to cover the range of possibilities, we consider i) individual oscillation modes becoming resonant with the tide, ii) the elliptical instability, where a pair of inertial modes exhibit a nonlinear resonance with the tide, and iii) the non-resonant p-g instability which may arise  as high order p- and g-modes in the star couple nonlinearly to the tide. In each case, we estimate the amount of additional energy loss that needs to be associated with the dynamical tide in order for the effect to impact on an observed gravitational-wave signal.  We explore to what extent the involved neutron star physics may be considered known and how one may be able to use observational data to constrain theory.
\end{abstract}

\maketitle

\section{Introduction}

With the announcement of the first direct detection of gravitational waves from the late inspiral phase of a double neutron star system \cite{bns} we  enter an exciting new era for gravity, astrophysics and nuclear physics. From the gravity perspective, binary merger signals lend support for Einstein's theory in the dynamical strong field regime. From the astrophysics point-of-view, one may anticipate that observed event rates will give insight into the formation channel(s) for these systems \cite{progen}. Moreover,  the positive identification of  electromagnetic counterparts to the merger events \cite{emc} should eventually confirm the theory paradigm for short gamma-ray bursts \cite{grb} and may shed light on the r-process nucleosynthesis (e.g. through a kilo-nova signature \cite{kilonova}). Finally, these observations are of tremendous importance for nuclear physics as they help unlock the secrets of the equation of state for matter at supranuclear densities. 

Neutron star binaries allow us to probe  equation of state physics in several unique ways, ranging from subtle to dramatic. As the stars enter the sensitivity band of ground-based detectors finite-size/fluid effects come into play. The interaction with the binary companion raises a tide of height
\beq
\epsilon \approx {M_2 \over M_1} \left( {R_1 \over a}\right)^3
\label{bulge}
\eeq
($M_1$ and $R_1$ are mass and radius of the primary, $M_2$ is the companion mass and $a$ is the binary separation) on each star. The response of the star, in the form of the tidal deformability encoded in the so-called Love numbers (which depend on the star's mass and radius) leaves a secular imprint on the gravitational-wave signal \cite{hind,read,agathos}. In addition, the star responds dynamically to the tidal interaction.  As the binary sweeps through the detector's sensitivity band resonances with various oscillation modes may become relevant. Notably, even though it  is not expected to exhibit a resonance before the stars merge, the tidal excitation of the star's fundamental f-mode is likely to be significant \cite{ftide}. This represents an aspect of the dynamical tide. The different tidal effects are expected to be subtle, but one would nevertheless hope to match observations to theory predictions to extract the stellar parameters  (mass and  radius for each of the two binary companions) and hence constrain the cold equation of state. 

The final merger provides a contrast in complexity. The violent merger dynamics requires full nonlinear simulations (see \cite{rez} for a recent review of the state of the art), implementing a challenging range of physics (from magnetohydrodynamics to neutrino transport). Nevertheless,  simulations \cite{baus,rez2,bern} hint at robust signal features which may eventually provide insight into thermal aspects of the equation of state. However,  detecting these features with the current generation of instruments is a serious challenge since the merger signal peaks at several kHz \cite{clark}. 

If we want to envisage a realistic scenario for the first set of binary neutron star  detections, then we can 
combine the fact that no signals were observed in the initial LIGO era with the sensitivity level of the first observation runs of the advanced instruments \cite{O1}. This suggests that one would  expect the first signals to cut off well below the merger frequency. Any extraction of nuclear physics information then relies on precise template matching for the inspiral phase. This should be within reach as long as the only relevant feature is the tidal compressibility. However, there are aspects of the problem that may not (at least not any time soon) be accessible to detailed modelling. The aim of this paper is to explore (some of) the relevant issues, raise awareness of the associated modelling challenges and outline an observation-led approach that may (eventually) help constrain theoretical parameters from a given signal. 

As an illustration of the range of possibilities, we consider three dynamical effects associated with a star's tidal response. First we revisit the problem of resonances, where a given stellar oscillation mode grows as it becomes resonant with the orbital frequency \cite{l94,ks}. 
Our second example is, as far as we are aware, novel for neutron stars: We estimate the role of the elliptical instability \cite{kers,lieb,ogilvie} on the orbital evolution. This  is also a resonant phenomenon, although in this case it is a pair of inertial modes that couple to the tide. The third, and final, example is provided by the so-called p-g instability \cite{wab,nw}. This is a non-resonant mechanism which is supposed to arise due to a strong nonlinear coupling between high-order pressure (p) and  gravity (g) modes in the star's core.  It has been suggested that this instability becomes active when the system evolves beyond 50~Hz or so, i.e. shortly after the signal enters the detector band, and that it grows to the point where it has severe impact on the gravitational-wave phasing \cite{ess}. Unfortunately, it is far from easy to establish to what extent this is a real concern as the p-g instability involves  short wavelength oscillation modes, and these are sensitive to the internal physics. 

In order to assess the relevance of dynamical effects associated with the tidal interaction, we will make use of a series of back-of-the-envelope level estimates. These may not be particularly useful if one is interested in precise statements, but they provide an immediate idea of issues that may warrant more detailed attention. We consider the energetics of additional mechanisms that may sap energy from the binary orbit, and hence impact on the gravitational-wave signal. As a simple measure, we focus on the number of wave cycles in a signal between a frequency $f=f_a$, when the signal first enters the detector band, and $f_b$, when it dives into the noise again. As a specific example, we will consider the (potentially conservative) frequency range  from $f_a\approx 30$~Hz to $f_b\approx 300$~Hz in the following.

The idea is simple; once an additional mechanism leads to an overall shift of about half a cycle  in the waveform then there would be no further accumulation of signal-to-noise in a matched filter search \cite{read}. Hence, if the total number of cycles is $\mathcal N=\mathcal N(f)$  (where $f$ is the gravitational wave frequency) then a shift $\Delta \mathcal N > 0.5$  (or equivalently, a phase shift $\Delta \Phi = 2\pi\Delta \mathcal N$ of order a few radians) would suggest that the effect could be distinguishable. This rough criterion will be sufficient for our  (somewhat qualitative) discussion. Basically, we assume that an additional dynamical effect would i) suppress detectability and affect parameter extraction with a given search template (that does does account for the effect) if $\Delta \mathcal N > 0.5$, but it should be safe to assume that, ii) the effect will not be distinguishable if $\Delta \mathcal N \ll 1$. 

The ultimate aim of the discussion is to consider two questions: Do we need to worry about ``unmodelled'' aspects of the tidal problem? If so, to what extent can we use observational data to constrain the involved theory?

\section{Inspiralling binaries}

We take the leading-order gravitational radiation reaction as our starting point. That is, we assume that gravitational-wave emission drains energy from the orbit at a rate 
\beq
\dot E_\mathrm{gw} = - {32 \mathcal M \Omega \over 5 c^5} \left( G \mathcal M \Omega\right)^{7/3}
\eeq
where the chirp mass is given by
\beq
\mathcal M = \mu^{3/5} M^{2/5} = M_1 \left ( {q^3 \over 1+q} \right)^{1/5}
\eeq
with the total mass $M=M_1+M_2$, reduced mass $\mu=M_1M_2/M$ and mass ratio $q=M_2/M_1$.  In the case of a pair of $1.4 M_\odot$ neutron stars (which we take as our canonical example throughout the discussion) we have $\mathcal M= 1.2M_\odot$. 

As we are considering how we can use observations to constrain the involved neutron star physics, it is important establish to what extent the various parameters are already known. For the mass ratio $q$, we know from radio observations that double neutron systems may be asymmetric, as in the case of PSR J0453+1559 where the two masses are $1.174M_\odot$ and $1.559M_\odot$ \cite{mart}. Given this, it would not be surprising to find a mass ratio in the range (taking the primary to be the heavier companion) $0.7\le q\le 1$.

The orbital frequency $\Omega$ follows from Kepler's law
\beq
\Omega^2 = {GM\over a^3} 
\eeq
which links the observed gravitational-wave frequency 
\beq
f= {\Omega \over \pi} 
\eeq
to the orbital separation $a$.

Given  the Newtonian orbital energy
\beq
E_\mathrm{orb} = E_\mathrm{N} = -  {GM_1 M_2 \over 2a}  = - {\mathcal M\over 2} \left( G\mathcal M\Omega\right)^{2/3}
\label{eorb}
\eeq
it follows that the orbit evolves in such a way that
\beq
{\dot \Omega \over \Omega}  = - {3\over 2} {\dot a \over a}
 = {3\over 2} {\dot E_\mathrm{orb} \over E_\mathrm{orb}} \approx  {3\over 2} {\dot E_\mathrm{gw} \over E_\mathrm{N}}
 =  {96 \over 5 c^5} ( G\mathcal M \Omega)^{5/3} \Omega \equiv {1\over t_D}
\label{Deqn}
\eeq
defines the inspiral timescale $t_D$. That is, we have
\beq
t_D \approx  140 
 \left({\mathcal M \over 1.2\,M_\odot}\right)^{-5/3}
 \left({f \over 30\mbox{ Hz}}\right)^{-8/3}\ \mathrm{s}
\eeq
The two neutron stars will merge about 2 minutes after the system enters our assumed frequency range.
The result also manifests the well-known fact that the leading order gravitational-wave signal only encodes the chirp mass. However, one would expect to be able to extract the individual masses (and possibly the spins) from higher order post-Newtonian corrections \cite{pNreview}.  
This is important as the stellar parameters enter the discussion of the tidal response. These effects are, of course, subtle and a key question concerns to what extent unmodelled features may limit the precision of the parameter extraction. It is important to keep in mind that, while one may expect to obtain fairly good estimates for the individual masses, it will be more difficult to infer the individual spin rates (the spin-spin and spin-orbit coupling effects are likely to be weak). 

As long as it is safe to ignore other aspects, the binary signal would be associated with a total number of cycles; 
\beq
\mathcal N_\mathrm{gw} = \int_{t_a}^{t_b} f dt = \int_{f_a}^{f_b} {f \over \dot f } df =   \int_{f_a}^{f_b} t_D df
= {c^5 \over 32 \pi \left( G \mathcal M \pi f_a \right)^{5/3} } \left[1 - \left( {f_a \over f_b}\right)^{5/3} \right]
\eeq
For our example frequency range the total number of cycles would be $\mathcal N_\mathrm{gw} \approx 2500$.

Let us now consider the possibility that the tidal dynamics leads to some additional change of orbital energy, say at a rate $\dot E_\mathrm{tide}$. This will lead to a change in the number of wave cycles in the observed frequency range. Specifically, with 
\beq
\dot E_\mathrm{orb} = \dot E_\mathrm{gw} + \dot E_\mathrm{tide}
\eeq
we have
\beq
\mathcal N =  {2\over 3} \int_{f_a}^{f_b} {E_\mathrm{orb} \over \dot E_\mathrm{orb}} df   
\approx  \int_{f_a}^{f_b} t_D  \left( 1 - {\dot E_\mathrm{tide} \over \dot E_\mathrm{gw} }  \right) df = \mathcal N_\mathrm{gw} + \Delta\mathcal N
\label{Napprox}
\eeq
where the last step should be a good approximation if $\dot E_\mathrm{tide} \ll \dot E_\mathrm{gw}$. We see that  the additional torque leads to a contribution; 
\beq
\Delta \mathcal N =  - \int_{f_a}^{f_b} t_D \left( {\dot  E_\mathrm{tide} \over \dot E_\mathrm{gw} }  \right) df \label{dN}
\eeq
This  allows us to estimate the relevance of any mechanism that is active through the observed frequency range. Note that, even though one might intuitively expect an increase in the rate of inspiral, e.g. a decrease in the number of cycles, there may be situations where the opposite happens and an additional mechanism pumps energy into the orbit. In this case the number of cycles would obviously increase. We discuss a particular example of this later. 

Moreover, we have not accounted for any changes to the orbital energy associated with the tidal effect. If we do this, say, by letting 
\beq
E_\mathrm{orb} = E_\mathrm{N} + E_\mathrm{r}
\eeq
then we arrive at
\beq
\mathcal N =  {2\over 3} \int_{f_a}^{f_b} {E_\mathrm{orb} \over \dot E_\mathrm{orb}} df   
\approx  \int_{f_a}^{f_b} t_D  \left( 1 + {E_\mathrm{r} \over E_\mathrm{N}} - {\dot E_\mathrm{tide} \over \dot E_\mathrm{gw} }  \right) df 
\label{Napprox2}
\eeq
As we will see later, the additional phase shift may be important, but it is not all that easy to quantify for the dynamical problems we will consider. To do this one would need an explicit model for the interaction between the tidally excited fluid motion and the orbital dynamics and this would inevitably involve unknown ``efficiency'' factors.

We also need to consider the possibility of resonances, where the additional energy loss is associated with a (more or less) distinct frequency. In this case, we may rewrite \eqref{dN} as 
\beq
\Delta \mathcal N \approx -   \int_{f_a}^{f_b} f {d E_\mathrm{tide} \over da} {1 \over \dot E_\mathrm{orb} }  da
\eeq
After integration, this leads to (cf. \cite{l94}, noting the different definition of $t_D$)
\beq
\Delta \mathcal  N \approx -  \left( {f \Delta E_\mathrm{tide} \over \dot E_\mathrm{orb} } \right)_{f=f_\alpha} \approx - \left( {3 \over 2} {f t_D \Delta E_\mathrm{tide} \over E_\mathrm{N} } \right)_{f=f_\alpha}
\label{dNres}
\eeq
where $\Delta E_\mathrm{tide}$ is the total energy transferred from the orbit to the resonant mode, and the expression should be evaluated at the resonance frequency $f=f_\alpha$   (where $\alpha$ is a label that identifies the resonant mode).

\begin{figure}[h]
\begin{center}
\includegraphics[width=10cm,clip]{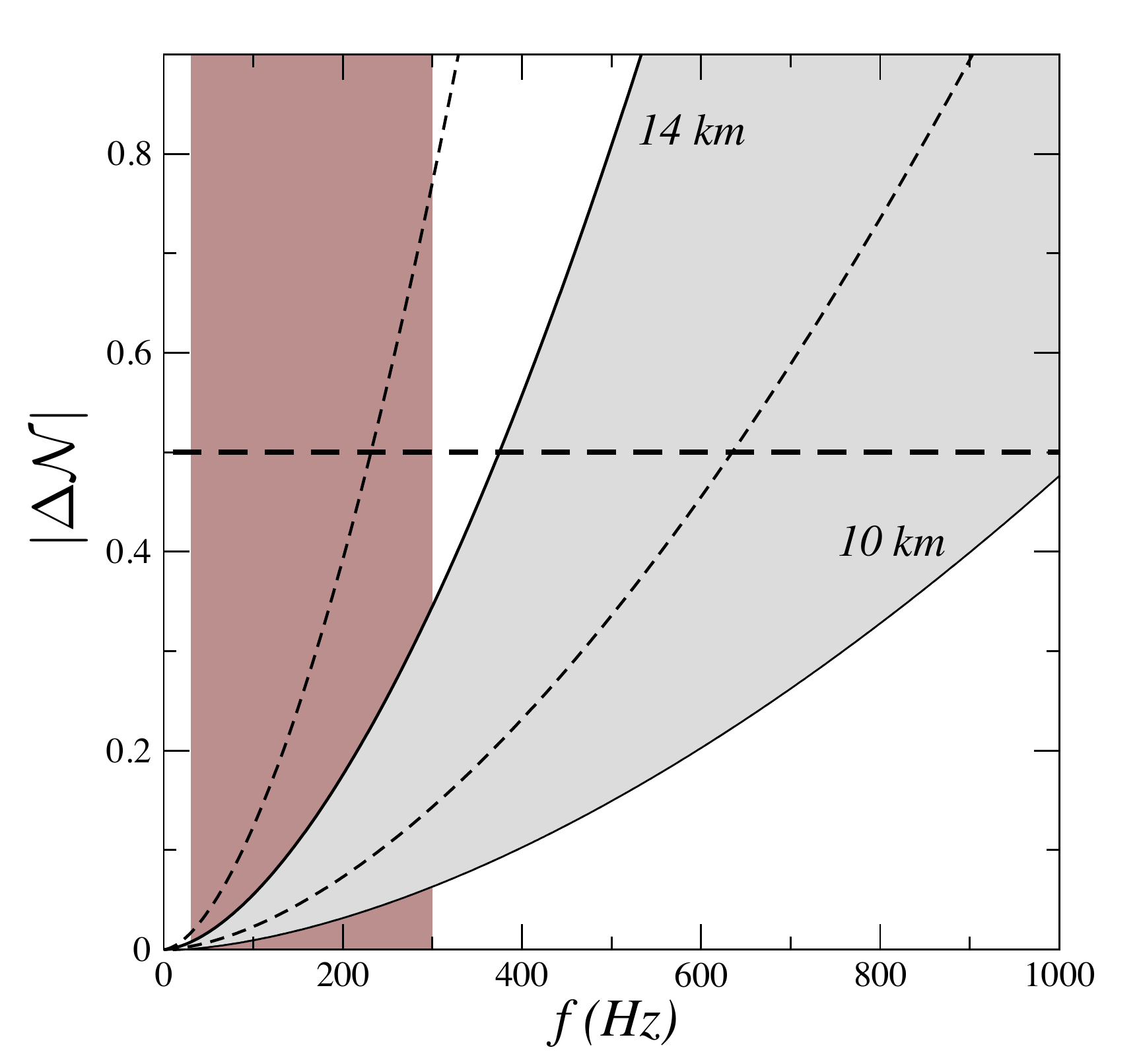}
\end{center}
\caption{A schematic illustration of the impact of tidal compressibility on a binary neutron star signal. We show the estimated shift in the number of gravitational wave cycles $|\Delta \mathcal N|$ as a function of the gravitational-wave frequency $f$. The grey band follows from \eqref{estlove} if we assume a Newtonian $n=1$ polytrope (for which $k_2\approx 0.26$), two equal $1.4M_\odot$ neutron stars (thus doubling the value of $\Delta \mathcal N$) and the  ``reasonable'' range of radii $10-14$~km. The dashed curves show how this band shifts if we consider the (likely unrealistic) case of two $1.1 M_\odot$ stars.  (For more detailed/realistic models, see for example Fig.~4 in \cite{read}.) The dashed horisontal line represents the indicative level of $|\Delta \mathcal N| \approx 0.5$ above which the effect will leave an imprint in a matched filter search, and the vertical shaded region represents our example frequency range between $30$ and $300$~Hz.} 
\label{Love}
\end{figure}

As a useful comparison and illustration of the level of uncertainty of the discussion, let us sketch the impact of the tidal compressibility.  
From the results of  \cite{read} we have the quadrupole contribution (from one of the stars)
\beq
2\pi \Delta \mathcal N \approx - {13\over 2} {1 \over q(1+q)^{4/3}} \left( {c^2 R_1 \over GM_1}\right)^{5/2} \left( {\pi f \over \Omega_0}\right)^{5/3} \tilde k_2
\label{estlove}
\eeq
with
\beq
 \Omega_0 = \left( {GM_1 \over R_1^3} \right)^{1/2} \approx 2\pi\times 2200\mbox{ Hz}\left({M_1 \over 1.4\,M_\odot}\right)^{1/2}\left({R_1\over 10\mbox{ km}}\right)^{-3/2}
\eeq
where $R_1$ is the radius of the neutron star and $\tilde k_2$ a weighted average of the Love number;
\beq
\tilde k_2 = {1\over 26} (1+12 q) k_2
\eeq
Typical results, for an equal mass binary of Newtonian $n=1$ polytropes (for which $k_2=0.26$), are shown in Figure~\ref{Love}. The figure illustrates that the tidal compressibility comes into play at late stages of inspiral. In the example provided in the figure the effect would not be ``detectable'' below $f\approx 400$~Hz, i.e. it would become relevant outside our chosen frequency range. Of course,  we should keep in mind that less massive neutron stars leave a stronger imprint on the signal \cite{read}, cf. Figure~\ref{Love}.

In order to arrive at \eqref{estlove} one has to, first of all, consider the (formally 5th order post-Newtonian contribution) contribution to the radiation reaction;
\beq
\dot E_\mathrm{tide} = \dot E_\mathrm{gw} {4(1+3q) \over (1+q)^{5/3}} \left( {\pi f\over \Omega_0}\right)^{10/3} k_2
\label{edotalt}
\eeq
and  the corresponding contribution to the orbital energy \cite{read}. If we were to account for only the first of these, e.g. strictly follow the strategy that led to \eqref{dN}, then the final result would  differ by about a factor of 2 for the equal mass case. In fact, the contribution to the orbital energy dominates. As we do not expect the estimates in the following to be accurate to factors of order unity this may not be a major concern. However, it is important to understand that this is the level of ``accuracy'' we are working at.  It is also worth noting that this factor would sufficiently decrease the impact of the tidal compressibility that it would only come into play at very high frequencies, cf. Figure~\ref{Love}. This demonstrates the importance of more precise modelling but, as we shall see,  dynamical tides involve physics that is sufficiently uncertain that this may not (yet) be within reach.

\section{Resonances}

Neutron stars have rich internal dynamics with different sets of oscillation modes, more or less directly associated with specific aspects of the physics \cite{krug}. The modes are typically characterised in terms of a harmonic time dependence $e^{i\omega t}$ and an expansion in spherical harmonics $Y_{lm}$, with $l$ associated with the polar angle and $m$ following from the dependence on the azimuthal angle as $e^{im\varphi}$. This decomposition is ``clean'' for non-rotating stars, but rotation couples the different harmonics, which makes a study of the seismology of fast spinning neutron star more challenging \cite{gaertig}. However, binary systems that enter the sensitivity band of ground based gravitational wave detectors will be relatively old, so it seems reasonable to assume that the stars would have had ample time to evolve and spin down to modest rotation rates. The fastest  known pulsar in a double neutron star system, PSR J0737-3039A, currently spins  with a period of $P =1/f_s= 0.0227$~s \cite{dpsr}. If we take the observed spin-down rate  $\dot P = 1.76\times 10^{-18}$ and assume  the standard magnetic dipole braking index of $n=3$  we can evolve this system into the distant future. We then find that this pulsar would enter the LIGO band spinning at a frequency $f_s \approx 35$~Hz. This may seem a high rate of spin, but it is far below the break-up speed (see below). At this rotation rate the star is nearly spherical so a (slow-rotation) expansion (in $Y_{lm}$) should  be adequate. This conclusion may, of course, change if a faster spinning object were to be found in a neutron star binary. In this context, it may be worth noting PSR J1807-2500B which spins at 239~Hz \cite{lynch}, but the nature of the binary partner in this system is uncertain. For evolutionary reasons, as it provides a natural explanation for the fast spin, one may expect the companion to be a white dwarf. 

The tidal interaction provides a driving force on the stellar fluid, characterised by the time dependence $e^{im'\Phi(t)}$ where $\Phi(t)\approx \Omega t$ as long as the evolution is adiabatic. The tidal driving is also expanded in harmonics and the general expression may lead to different couplings to the fluid motion, especially if the orbit is eccentric. However, as we are focussing on the main features, we will consider the simplest case of a circular and coplanar orbit  in the following. In this case, the main tidal effect arises from the quadrupole $l=2$ coupling, and we also have $m=m'$. We are then left with two distinct components. The equilibrium tide is associated with $m=0$ while the dynamical tide arises from $m=\pm2$. In order to simplify the discussion we will also focus in the tidal response of one of the stars. A realistic signal discussion would have to  combine the results for the two stars (see for example \cite{read}), but as we are not aiming to make particularly precise statement we leave this for future work.

\subsection{Non-rotating stars}

Let us first consider the case of a non-rotating star, in which case the different $m$ harmonics are degenerate. In this case, we can use the results from \cite{l94};
\beq
\Delta E_\mathrm{tide} \approx - {\pi^2 \over 512} \left( {GM_1^2 \over R_1}\right) \hat \omega_\alpha^{1/3} Q_\alpha^2 \left( {R_1 c^2 \over GM_1} \right)^{5/2} q \left( {2 \over 1+q}\right)^{5/3}
\label{det}
\eeq
where we have introduced the dimensionless mode frequency $\hat \omega_\alpha$ through
\beq
\omega_\alpha = \hat \omega_\alpha \Omega_0
\eeq
The remaining parameter,  $Q_\alpha$, encodes the ``overlap integral'' which determines the strength of the tidal coupling to particular stellar oscillation modes. We also have the resonance condition
\beq
\omega_\alpha = 2\pi f_\alpha = 2\Omega =  2\pi f
\eeq
It is important to note that, for the quadrupole case, the oscillation frequency of the resonant mode ($f_\alpha$) is equal to the observed gravitational-wave frequency ($f$).

If we introduce the resonance radius
\beq
a_\alpha = \left[ {4GM_1(1+q)\over \omega_\alpha^2} \right]^{1/3}
\eeq
it readily follows that
\beq
E_\mathrm{N} = - {1\over 2^{5/3}} \left( {GM_1^2 \over R_1} \right) \hat \omega_\alpha^{2/3} { q \over (1+q)^{1/3}}
\eeq
and
\beq
{\Delta E_\mathrm{tide} \over E_\mathrm{N} } \approx  {\pi^2 \over 128\times 2^{1/3}}(\pi {\hat f_\alpha})^{-1/3} Q_\alpha^2 \left( {R_1 c^2 \over GM_1} \right)^{5/2} \left( {2 \over 1+q } \right)^{4/3}
\eeq
where $\hat f_\alpha = \hat \omega_\alpha/2\pi$. At resonance, we also have 
\beq
f t_D = {5 \over 96 \pi} (\pi \hat f_\alpha)^{-5/3} \left( {c^2 R_1 \over GM_1}\right)^{5/2} { (1+q)^{1/3} \over q } 
\eeq
and it follows from \eqref{dNres} that
\beq
\Delta \mathcal N \approx  - 4\times10^{-4} \hat f_\alpha^{-2} Q_\alpha^2  \left( {c^2 R_1 \over GM_1}\right)^5 {1\over q (1+q)}
\label{dNnew}
\eeq
As one might have expected, this is likely to be a small effect. Still, it is instructive to consider to what extent the different contributions can be considered known. 
We have already discussed the expected range for the mass ratio $q$ (from radio observations). The star's compactness is also (although less so) constrained by observations. From x-ray observations of accreting neutron stars one would expect the radius of a $1.4M_\odot$ star to lie in the range $10-14$~km \cite{steiner} (we will take the lower end of this range as our canonical case in the following). As the mass-radius curve tends to rise steeply in the relevant mass range (for a typical equation of state) we might assume the radius to be inside this range for all plausible masses in a binary. (Note that this argument does not account for the softening effect of possible internal phase transitions.) This would constrain the compactness to the range
\beq
0.12  \le   { GM_1 \over c^2 R_1}  \le 0.24
\eeq
This introduces an uncertainty of about a factor of 30 in the above estimate for $\Delta \mathcal N$, illustrating the importance of obtaining tighter constraints on the neutron star radius. This is, of course, one of the main targets of the observations in the first place. One may hope to (eventually) get a tighter radius constraint from the tidal compressibility. In addition, a measurement of the neutron star radius to within 5\%  is a key science aim of the NICER mission which is currently flying on the International Space Station \cite{nicer}. 

\begin{figure}[h]
\begin{center}
\includegraphics[width=10cm,clip]{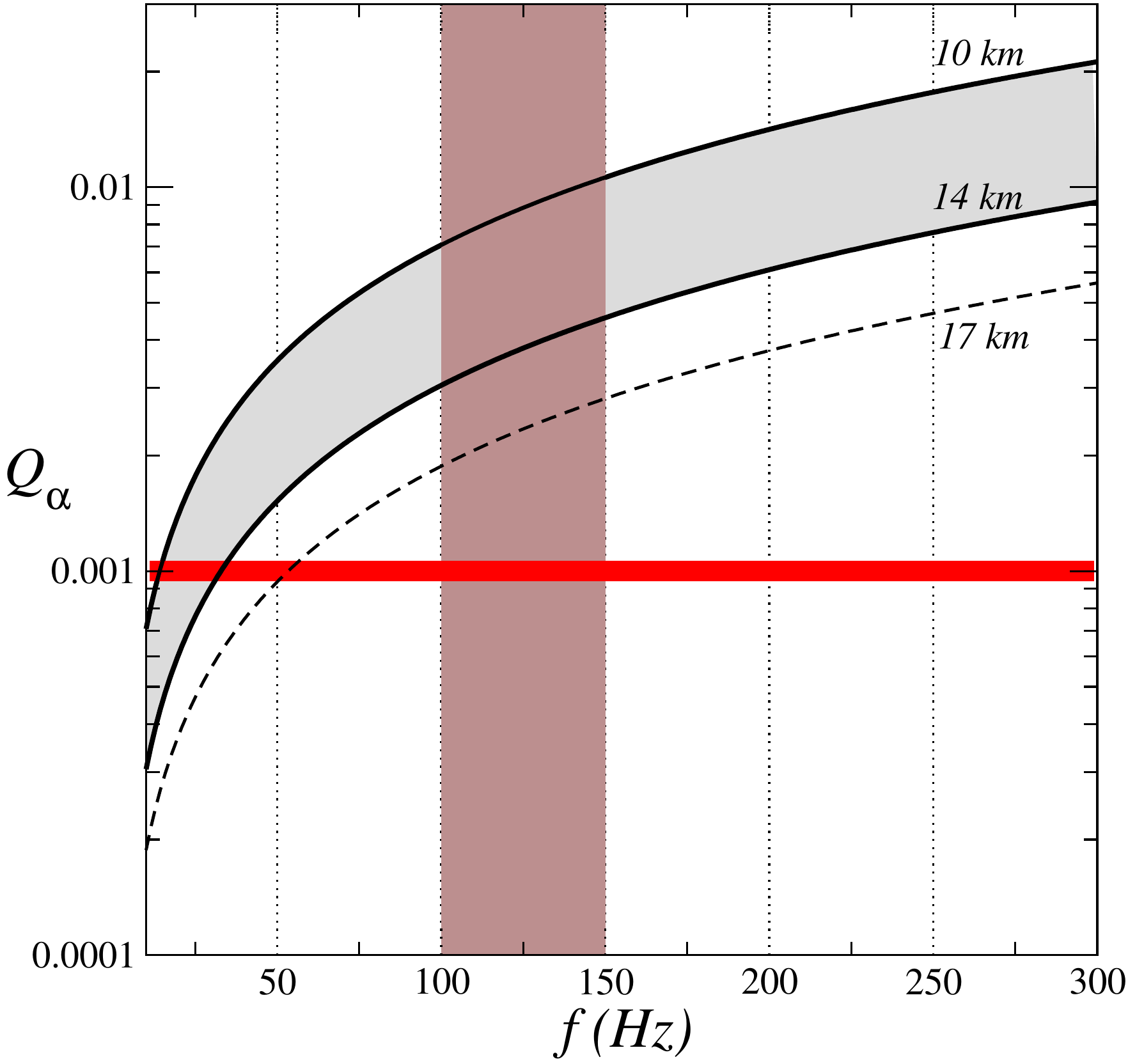}
\end{center}
\caption{ Constraints on $Q_\alpha$ if a limit $|\Delta \mathcal N|\le 0.5$ were to be inferred from inspiral data. The thin black lines represent equal mass $1.4M_\odot$ binaries with neutron star radius 10~km (upper curve) and 14~km (lower curve). The grey region represents the expected radius range from x-ray observations \cite{steiner}. As an indication, the thick horisontal (red)  line represents the largest values of $Q_n$ for the g-modes of a non-rotating star from \cite{l94}. This should be taken as indicative of what is expected from theory (with the caveats discussed in the main text). Finally, the shaded vertical region relates to an example where the observational constraint is obtained for a distinct frequency band (here taken to be 100-150~Hz). This figure illustrates that the resonant modes of a non-rotating star may be difficult to detect, but there could be a relevant effect below 50~Hz or so, if the neutron star radius were to be surprisingly large (the dashed curve shows the result for a radius of 17~km). One should also keep in mind that rotation may lead to slightly larger values of $Q_\alpha$, in which case the chance of detection would improve.} 
\label{resonance}
\end{figure}

With a narrower region of uncertainty for the stellar compactness, one may be able to use observed deviations from a pure radiation reaction inspiral to constrain the value of  $Q_\alpha$ for any resonant mode in a given frequency range. We illustrate this idea in Figure~\ref{resonance}. 
Imagine that one sets an upper limit on the deviation from a post-Newtonian radiation reaction inspiral  of order $\Delta \mathcal N \le 0.5$ in a given  frequency range, say $f = 100-150$~Hz. Then, we know from \eqref{dNnew} (assuming canonical neutron star parameters) that 
\beq
Q_\alpha \le 10^{-2} \left( {f \over 100~\mathrm{Hz}}\right) |\Delta \mathcal N|^{1/2}
\eeq
 This constraint is shown in Figure~\ref{resonance}. We see that the chances of observing the imprint of a tidal resonance is better at frequencies below a few tens of Hz. Moreover, given the dependence on the stellar compactness, the effect would be more prominent if the neutron star radius is large. In fact, given a reliable theoretical calculation for $Q_\alpha$ one can turn this argument into a constraint on the stellar radius.

In order to understand the wider implications of this kind of constraint for neutron star physics, we need to consider the nature of  specific oscillation modes. 
For non-rotating stars, the most likely set of modes to exhibit tidal resonance are the gravity g-modes. In a mature (cold) neutron star, these modes are associated with internal composition stratification \cite{reis}. If the motion of a moving fluid element is faster than the nuclear reactions that would equilibrate the fluid to its new surroundings, then the chemical differences lead to a buoyancy that provides the restoring force for these modes. In  the simplest models, the g-modes are associated with the varying proton fraction. This typically leads to mode frequencies below a few 100~Hz and a dense spectrum of high overtone modes at lower frequencies (see \cite{krug} for the current state of the art). The lowest order  (highest frequency) mode couples the strongest to the tide, with a typical value of the coupling constant $Q_n\approx 10^{-4}-10^{-3}$ \cite{l94}. Most likely, this makes the effect too weak to be detected by the current generation of instruments, see Figure~\ref{resonance}.

The discussion is nevertheless interesting.
The g-modes rely on physics beyond the bulk properties of the star, reflecting how the strong interaction determines the composition of matter at high densities. The state of matter is also important. For example, if the star's core contains a  superfluid then the charged components (in the simplest case, protons and electrons) can move relative to the neutrons. As a result, as long as we assume that the electrons and protons are electromagnetically coupled, the origin of the buoyancy is removed and there will no longer be any g-modes \cite{nac}. This would obviously remove any related resonances. However, there are twists to this story. The composition of a neutron star core is more complex than pure neutron-proton-electron matter. Close to the nuclear saturation density  the formation of muons becomes energetically favourable. This leads to stratification (now associated with the electron-muon fraction) also in a superfluid star \cite{gk,pass} which reinstates the composition g-modes. These new g-modes are expected to have  higher frequencies, perhaps by a factor of a few, which means that resonances become relevant at later stages of the inspiral. The first estimates of the tidal coupling for these modes \cite{yw} suggest that they may be associated with an increased transfer of energy but this is compensated for by the fact that the inspiral is accelerated at the higher frequencies. As a result, the estimated values of $Q_n$ are similar to those of the original g-modes  in \cite{l94}. The example in Figure~\ref{resonance} then suggests that the higher frequency g-mode resonances are likely to leave a weaker imprint on the gravitational-wave signal.

\subsection{Rotating stars}

The resonance problem becomes more intricate for rotating stars. First of all,  we need to note that the resonance condition  involves the mode frequency in the inertial frame. That is, we have (again for the quadrupole tide)
\beq
\omega_\alpha^{(i)} =  2\Omega 
\eeq
Secondly, rotation breaks the degeneracy associated with the azimuthal angle and modes associated with different values of $m$ become distinct (i.e. we need to separately consider $m=\pm2$). To leading order, the mode frequency is then given by 
\beq
\omega^{(i)}_\alpha = \omega^{(r)}_\alpha - m\Omega_s
\label{freqshift}
\eeq
where $\omega^{(r)}_\alpha$ is the frequency of the mode in the rotating frame and $\Omega_s$ is the spin of the star. If the star spins rapidly then the change in shape due to the centrifugal force provides an additional correction, but (as we have already suggested) it seems reasonable to argue that most binary systems will be old enough that the neutron stars would have slowed down significantly by the time they become detectable with ground based interferometers. The relevant comparison is the break-up rate, well approximated by 
\beq
\Omega_K \approx {2\over 3} \left( \pi G \bar \rho \right)^{1/2} ={ 1  \over \sqrt{3}}  \Omega_0 
\eeq
where $\bar \rho$ is the average density. Clearly, we have $\Omega_s/\Omega_K \ll1$ even for  the fastest known pulsar (PSR J1748-2446ad at 716~Hz \cite{wessels}).   Hence, we will not consider the quadratic rotational shape corrections. Nevertheless, it is clear from \eqref{freqshift} that low-frequency modes may be significantly affected by the rotation.

The obvious fact that rotation adds parameters to the problem (and that the individual spin rates may be difficult to extract from the inspiral waveform) complicates any effort to use an observed signal to constrain the physics. Schematically  we need to replace \eqref{det} by \cite{holai}
\beq
\Delta E_\mathrm{tide} \sim {\omega_\alpha \over \omega_\alpha - T_\alpha} \left( W \mathcal D Q\right)^2 
\label{rotde}
\eeq
where $T_n$ encodes the effect of the Coriolis force on the fluid motion, $W$ is a numerical factor arising from the spherical harmonics expansion of the tidal field 
and $\mathcal D$ (the Wigner function) is another factor associated with the rotation of the coordinate system to an axis orthogonal to the orbital plane. These numerical factors are, in principle, known once it is established which specific mode (and harmonic) is being considered, but from the observational point-of-view they are (most likely) part of the unknowns. The main point is that one would have to disentangle the different contributions in order to make a clear match with theory and this is not going to be straightforward. Nevertheless, it is worth noting qualitative features. In particular, if there is a near cancellation, the denominator in \eqref{rotde} could lead to an enhanced impact on $\Delta \mathcal N$. For specific rotation rates one may find that the impact of particular modes stands out. However, according to the results of \cite{holai} the overall phase shift will still be below $0.01$.  The potential impact of this enhancement is clear from Figure~\ref{resonance}.

The rotation-induced shift of the mode frequency \eqref{freqshift} may also make a given mode susceptible to the Chandrasekhar-Friedman-Schutz instability, where the oscillation is driven unstable by the emission of gravitational waves \cite{fs78}. The instability sets in when a retrograde mode in the rotating frame becomes prograde in the inertial frame. Effectively, the mode energy then becomes negative. In an isolated star one would not expect  the instability of g-modes to be particularly relevant because these modes are not efficiently emitting gravitational waves and the instability does not overcome viscous damping \cite{lai99}. The case of tidal driving is different. The main impact of a mode being formally unstable is that the growth of the  mode pumps energy back into the orbit. This would alter the sign of $\dot E_\mathrm{tide}$, slow the down the inspiral and lead to an increase in $\mathcal N$ rather than a decrease. This is an important feature, which if observed would provide exciting insight into the stellar dynamics. 

A spinning star also has a richer spectrum of oscillation modes. The Coriolis force provides an additional restoring force, which brings new sets of modes into existence \cite{lf} These inertial modes, quite naturally, scale with the rotation frequency. This means that one may easily confuse the identification of an observed resonance. However, if the inspiral signal provides an independent constraint on the star's spin then one could potentially rule out inertial modes as they have to lie in the range $-2\Omega_s \le \omega^{(r)}_\alpha \le 2\Omega_s$. The first estimates of tidal inertial-mode excitation \cite{holai} suggested that the impact would be minor, but a more recent analysis \cite{rac} demonstrated that the gravito-magnetic coupling  enhances the importance of the inertial modes. In the specific case of the r-modes, the phase shift may amount to \cite{rac} \beq
\Delta \mathcal N \approx {0.1\over 2\pi} \left( {R_1 \over 10\ \mathrm{km}}\right)^4 \left( {M_1 \over 1.4 M_\odot}\right)^{-10/3} \left( {f_s \over 100\ \mathrm{Hz}}\right)^{2/3}
\label{rmode}
\eeq
where $f_s = \Omega_\s/2\pi$. Based on the available results, this may be the strongest relevant mode resonance. As the r-mode is generically unstable to gravitational-wave emission \cite{na98}, the main resonance effect would then tend to slow down the inspiral.  Moreover, as the r-mode frequency is (not accounting for relativistic effects) given by $f = 4f_s/3$, an observational constraint on the star's spin would directly indicate the frequency of the associated resonance.

Finally, it is important to appreciate the competing nature of the different restoring forces. Modes that are dominated by buoyancy in a non-rotating star may become  inertial above some rotation rate (when the Coriolis force becomes dominant) \cite{passg,kokk}. Such mixed inertia-gravity modes have only very recently been considered in the tidal context \cite{xu}.

\section{The elliptical instability}

As we have seen, the gravitational-wave signal from a neutron star binary may exhibit features due to individual oscillation modes becoming resonant. The resonance problem is conceptually straightforward, although it obviously involves poorly understood aspects of neutron star physics. However, the tidal problem may  be more subtle. In addition to individual mode resonances we may have to consider how nonlinearly coupled modes interact with the tide. An example of this is the elliptical instability. This is a parametric instability associated with the fact  that fluids that flow along elliptical flowlines tend to be unstable \cite{kers,lieb,ogilvie}. In essence, the instability acts through the nonlinear coupling of two inertial modes to the equilibrium tide. Recent work argues that the mechanism may be relevant for hot Jupiters \cite{bl1,bl2}, but the problem has not (as far as we know) been considered for binary neutron stars. Yet it is clearly relevant, at least conceptually. The tide raised by a binary companion deforms a neutron star and if  the star is spinning the internal fluid will flow along elliptical flow lines. The elliptical instability may come into play.

\subsection{Estimated timescales}

In order to explore the elliptical instability for binary neutron stars, we make use of order of magnitude estimates from the literature. First of all, we note that the instability is associated with inertial modes of the rotating star. These modes are confined to the frequency range \cite{lf}
\beq
- 2\Omega_s < \omega_\alpha^{(r)} < 2 \Omega_s
\label{inert}
\eeq
We can also think of the tidal bulge as a wave moving around the star \cite{ogilvie,bl1}. In the quadrupole case this would correspond to a rotating frame frequency
\beq
\omega_\mathrm{tide}^{(r)} = 2 (\Omega-\Omega_s)
\eeq
Now consider a pair of inertial modes with frequency $\mp \omega_0$ that interact nonlinearly with the tide. This interaction gives rise to disturbances with frequency
\beq
\omega_{nl} = \mp \omega_0 \pm \omega_\mathrm{tide}^{(r)}
\eeq
We see that, if $ \omega_\mathrm{tide}^{(r)} = 2 \omega_0$ then the nonlinear coupling amplifies the two original modes. This leads to the elliptical instability. 

This simple argument provides us with the necessary condition for the instability. Since the original modes have to lie in the inertial range \eqref{inert} we must have 
\beq
\Omega_s >  {\Omega \over 3} \qquad \mbox{or} \quad -\Omega_s < \Omega
\label{nec1}
\eeq
We will focus on the first case (co-rotation) from now on.

Once we establish that the instability may be active, we need an estimate of the growth rate. From \cite{bl1} we learn that the maximum growth rate of the instability is given by 
\beq
{1 \over t_\mathrm{el}} \approx {9\over 16} \epsilon |\Omega_s - \Omega|
\eeq 
where the tidal bulge, $\epsilon$, follows from \eqref{bulge}. Making use of Kepler's law, we see that 
\beq
\epsilon \approx { q \over 1 + q} \Omega^2 \left( {R_1^3 \over GM_1} \right)  = {1\over 3}{ q \over 1 + q} \left( {\Omega \over \Omega_K}\right)^2
\eeq
where $q$ is the mass ratio (as before). 
Thus, the final estimate for the growth timescale is
\beq
{1 \over t_\mathrm{el}} \approx {3\over 16} { q \over 1 + q} \left( {\Omega \over \Omega_K}\right)^2 |\Omega_s - \Omega|
\eeq
Finally, at the onset of instability $\Omega_s =  \Omega/3$, so let us (for simplicity) assume that $\Omega_s \gg \Omega$. Then we have
\beq
{1 \over t_\mathrm{el}} \approx {3 \over 16} { q \over 1 + q}  \left( {\Omega \over \Omega_K}\right)^2\Omega_s 
 \approx 0.2  { q \over 1 + q} \left( {f  \over  100\ \mathrm{Hz}} \right)^2 \left( {f_s \over 100\ \mathrm{Hz}}\right) \ \mathrm{s}^{-1}
\label{tg2}
\eeq

In order to establish whether the elliptical instability can be relevant for  inspiralling neutron star binaries, we compare the growth time to the  orbital evolution. 
For the instability to grow fast enough, we need $t_D\gg t_\mathrm{el}$. Combining the estimated timescales, we have 
\beq
{t_D \over t_\mathrm{el}} \approx 2.4\times10^{-2} {1\over (1+q)^{2/3}} \left( {R_1 c^2 \over GM_1}\right)^{5/2} \left({ \Omega_s \over \Omega_K}\right)  \left( {\Omega \over \Omega_K}\right)^{-2/3} \gg 1
\label{gt}
\eeq

 In the case of equal mass (canonical $M_1= 1.4M_\odot$ and $R_1=10$~km) neutron stars, we require
\beq
{t_D \over t_\mathrm{el}} \approx 0.54 \left(  {f_s \over 100~\mathrm{Hz}}\right) \left( {f \over 100~\mathrm{Hz}}\right)^{-2/3} \gg 1
\label{gt3}
\eeq
That is, the elliptical instability would grow fast enough to be ``interesting'' when
\beq
f_s \gg 185 \left( {f\over 100\ \mathrm{Hz}}\right)^{2/3}\ \mathrm{Hz}
\label{con1}
\eeq

\begin{figure}[h]
\begin{center}
\includegraphics[width=10cm,clip]{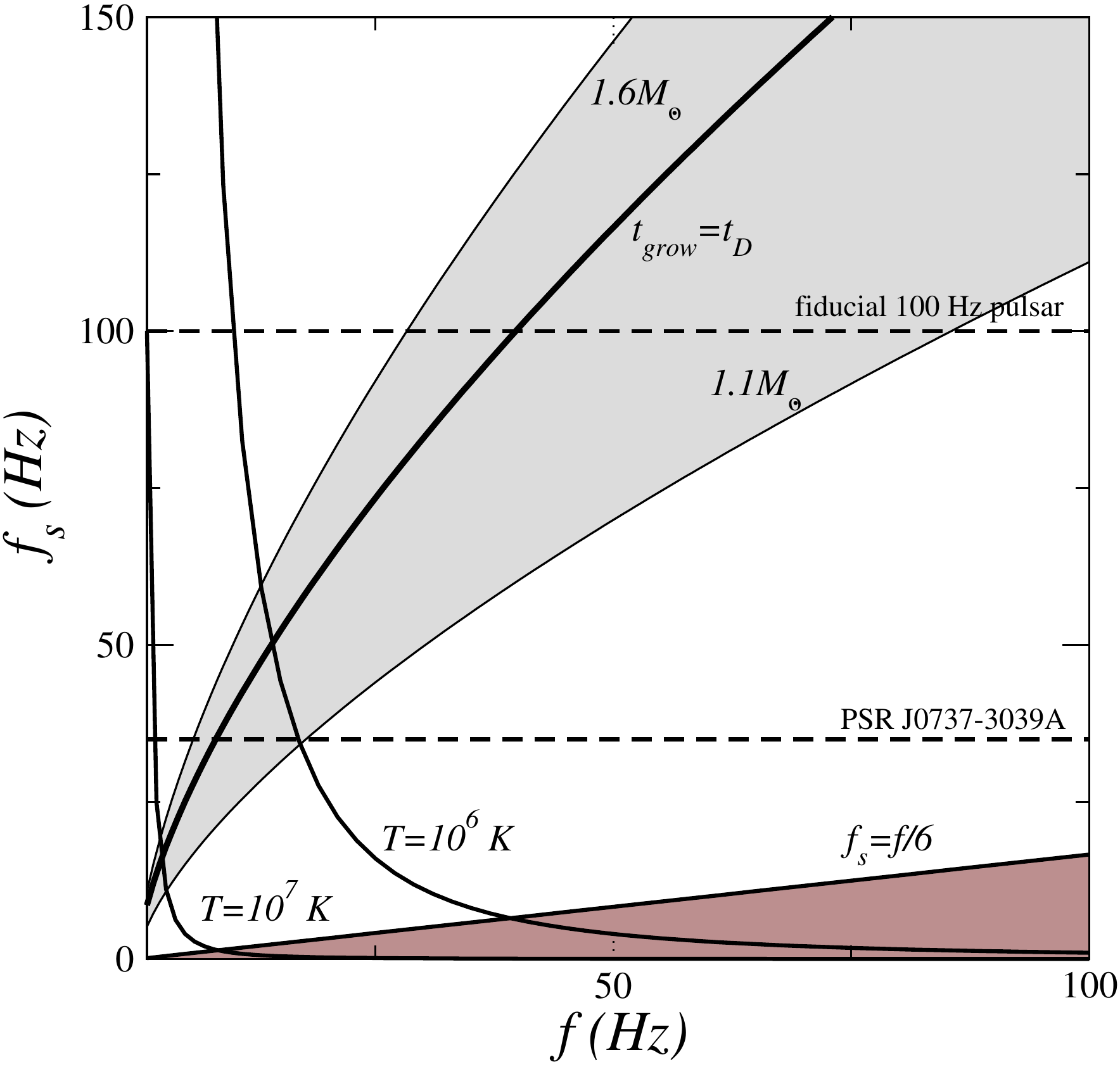}
\end{center}
\caption{Comparing the different frequency constraints to see if the elliptical instability may play a role for neutron star binaries. The vertical axis is the spin frequency, $f_s$,  while the horizontal one is the gravitational-wave frequency, $f$. The lower horizontal dashed line represents PSR J0737-3039A (extrapolated to be at 35 Hz as it enters the LIGO sensitivity band). A system has to be above all the other curves, arising from \eqref{con1}, \eqref{con2} and \eqref{con3} (showing both $T=10^6$~K and $10^7$~K), in order to be unstable. These estimates suggest that a system like PSR J0737-3039A may (briefly) exhibit the elliptical instability during the  inspiral phase. Any (at the moment fiducial) faster spinning pulsar could be more significantly affected.}
\label{freqs}
\end{figure}

Is this inequality likely to be satisfied for real systems? As already mentioned, we can extrapolate the evolution for PSR J0737-3039A to the point where the system enters the LIGO band. The pulsar would then spin at $f_s \approx 35$~Hz. In this case there may be a narrow frequency range where the elliptical instability can grow, see figure~\ref{freqs}. As yet unknown systems that  enter the detector band spinning at a faster rate may obviously be more significantly affected by the instability. 

Of course, we  need to keep in mind the necessary criterion \eqref{nec1};
\beq
f_s > {1\over 6} f\approx 17 \left( { f\over 100\ \mathrm{Hz}} \right) \ \mathrm{Hz}
\label{con2}
\eeq
This immediately shows that we need the neutron star to spin faster than 5~Hz or so for the formal instability regime to overlap our assumed signal frequency range (above $f=30$~Hz). One would expect many systems to satisfy this condition.

We also need to consider the impact of (shear) viscous damping. As we do not have the velocity fields for the unstable modes, and our discussion is at the back-of-the-envelope level anyway, let us progress by means of a rough estimate (see \cite{cl87} for a similar argument). The kinetic energy of an oscillation mode follows from
\beq
E_k = {1\over 2} \int \rho |\delta v|^2 dV
\eeq
where $\delta v$ is the velocity perturbation,
while the damping due to shear viscosity is given by
\beq
\dot E \approx - 2 \int \eta |\delta \sigma|^2 dV
\eeq 
where $\eta$ is the shear viscosity coefficient and $\delta\sigma\approx \nabla \delta v$ is the shear associated with the perturbation. Replacing the derivative with a characteristic lengthscale $L$ of a given oscillation, we have the timescale
\beq
t_\mathrm{sv} \approx - {2E_k\over \dot E} \approx {\rho L^2 \over 2 \eta}
\eeq
For low order modes the angular derivative dominates and we have $L\approx \pi R_1/l$ where $l$ is the usual spherical harmonics index. Meanwhile, for high overtones the radial derivative dominates so we have $L\approx R_1/n$ where $n$ is the overtone index of the mode.
The upshot of this is that the viscous damping timescale is estimated as 
\beq
t_\mathrm{sv} \approx \mathrm{min} \left[ {1\over n^2} , {\pi^2\over l^2}\right] {\rho R_1^2 \over 2\eta} \approx  \left[ {1\over n^2} , {\pi^2\over l^2}\right] {3 \over 8\pi} { M_1  \over  R_1 \eta}
\label{shear}
\eeq
where we have assumed a uniform density to arrive at the last estimate. This result does not differ too much from  precise calculations for 
specific oscillation modes so we expect it to provide a good indication of the relevance of viscosity. 

First of all, we can use \eqref{shear} to estimate when the viscosity wipes out the elliptical instability. To do this, we need the viscosity to dominate the instability growth up to the frequency given by \eqref{con2}, when the instability shuts down anyway. In a cold (superfluid) neutron star, the most important damping  is likely to be due to (electron-electron scattering) shear viscosity for which we have \cite{shear}
\beq
\eta \approx 2\times 10^{18} {\rho_{15}^{9/4} \over T_9^2}\ \mathrm{g/cm\ s}
\eeq
As we are considering low order inertial modes  coupled to the equilibrium tide, we take $l=2$  and arrive at the order of magnitude estimate 
\beq
t_\mathrm{sv} \approx 10^3 \left({M_1\over 1.4\,M_\odot}\right)^{-5/4}
\left({R_1\over 10\mbox{ km}}\right)^{13/4} \left( {T \over 10^6 \ \mathrm{K}}\right)^2 \ \mathrm{s}
\eeq
Combining this  
 with \eqref{tg2} we see that, for an equal mass (canonical neutron star) system we need
\beq
{t_\mathrm{sv} \over t_\mathrm{el} } \approx 100   \left( {f  \over  100\ \mathrm{Hz}} \right)^2 \left( {f_s \over 100\ \mathrm{Hz}}\right)  \left( {T \over 10^6 \ \mathrm{K}}\right)^2 >1
\eeq
in order for the instability to overcome viscosity.
This leads to the condition
\beq
f_s >   \left( {f\over  100\ \mathrm{Hz}} \right)^{-2}  \left( {T \over 10^6 \ \mathrm{K}}\right)^{-2}\ \mathrm{Hz}
\label{con3}
\eeq
This constraint is also illustrated in Figure~\ref{freqs}.

\subsection{Impact on the gravitational-wave signal}

Having discussed the conditions under which the elliptical instability may be active in a neutron star binary, let us estimate the effect it may have. First of all, let us consider the impact on the spinning star. In principle, \cite{ogilvie,bl1,bl2} the instability would tend to i) circularise the binary and ii) synchronise the spin of the star with the orbit. However, gravitational radiation reaction already leads to the orbital ellipticity evolving in such a way that binaries are expected to be circular by the time the signal enters the sensitivity band of a groundbased detector. Given this, we only consider the effect on the star's spin. 

The torque exerted on a spinning star by its companion is \cite{gold}
\beq
\dot J  = {9\over 4} {G R_1^5 \over D} {M_2^2 \over a^6} 
\eeq
where  $J$ is the angular momentum of the star and the dissipation function $D$ encodes the energy released per orbit,  the main unknown parameter in the problem. Introducing
the moment of inertia
\beq
I = \tilde I M_1 R_1^2
\eeq
where $\tilde I \approx 0.261$ for an $n=1$ polytrope  \cite{owen}, we see that the star's rotation frequency evolves in such a way that
\beq
\dot \Omega_s \approx - {3\over 4}{ 1\over \tilde I D} \left( {\Omega\over \Omega_K} \right)^2 \left( {M_2 \over M} \right)^2 \Omega^2 
\eeq
Next, we use the results of \cite{bl2} to estimate the dissipation function. Thus, we have
\beq
D \approx 9 \chi^{-1}  \left( {M\over M_1} \right) \left({ \Omega_K \over \Omega} \right)^4
\eeq
were $\chi$ is unknown, but the results from \cite{bl1,bl2} suggest that it might be reasonable to take $\chi \approx 10^{-2}$.
Combining the estimates, we have
\beq
\dot \Omega_s \approx - {\chi \over 12 \tilde I} \left( {M_1 M_2^2 \over M^3} \right) \left( {\Omega \over \Omega_K}\right)^6 \Omega^2  
\eeq
This provides us with an estimate of the synchronisation timescale;
\beq
t_s \approx {\Omega_s \over | \dot \Omega_s |} \approx {12 \tilde I \over \chi } \left( {M^3 \over M_1M_2^2} \right) \left( {\Omega_K \over \Omega}\right)^6 {\Omega_s \over \Omega} {1\over \Omega}  
\eeq
It follows that  (for an equal $1.4M_\odot$ system)
\beq
{t_s \over t_D} \approx 3\times10^7 \left( {f_s\over 100\ \mathrm{Hz}}\right) \left( {f \over 100\ \mathrm{Hz}}\right)^{-16/3} \gg 1
\eeq
Perhaps not surprisingly, the elliptical instability will not be able to synchronise the spin of the star to the orbit. In fact, the effect is tiny. As an illustration. let us consider the associated shift in the number gravitational-wave cycles. Assuming that the energy change associated with the torque is lost from the orbital energy, we use
\beq
\dot E_\mathrm{tide} \approx - \tilde I M_1 R_1^2 \Omega_s \dot \Omega_ s 
\eeq
in \eqref{dN}. This leads to 
\beq
{\dot E_\mathrm{tide} \over \dot E_\mathrm{gw}} \approx  0.2 { \tilde I \chi \over (1+q)^{7/3}} \left({ c^2 R_1 \over GM_1}\right)^{5/2}
\left( {\Omega_s \over \Omega_K} \right)   \left( {\pi f \over \Omega_0}\right)^{14/3} 
\eeq
In the specific case of an equal mass (canonical) binary and the assumed observed frequency range (ignoring the fact that the instability is unlikely to be active through the entire range) we have 
\beq
\Delta \mathcal N \approx - 5\times10^{-6} \chi \left( {f_s \over 100\ \mathrm{Hz}}\right)
\eeq
It would seem safe to ignore (remember that $\chi$ is a small number) the impact  the elliptical instability will have on the gravitational-wave signal. The mechanism is conceptually interesting, but unlikely to leave a detectable imprint. Nevertheless, the discussion  serves as a reminder that pairs of modes may couple nonlinearly to the tide. As it makes the problem significantly more complex, it is important to keep the possibility in mind. 

\subsection{Other binary systems}

The conditions for the elliptical instability to operate are more favourable if the neutron star spins rapidly. Hence, it makes sense to consider if it could play a role in accreting neutron stars in low-mass x-ray binaries, progenitors to the observed millisecond radio pulsars. In these systems, the neutron star spins at several 100 Hz and is accompanied by a low-mass partner so the typical mass ratio will be something like $q \approx 0.1$. Meanwhile, the orbital period tends to be several hours. For example, the fastest known accreting rotator, 4U1608-522, spins at 620 Hz and has an orbital period of 12 hours \cite{4Uref}. Scaling \eqref{tg2} to typical parameter values we have
\beq
 t_\mathrm{el} \approx 66,000  \left( {2\pi/\Omega \over 1\ \mathrm{hour}} \right)^2 \left( {f_s \over 500\ \mathrm{Hz}}\right)^{-1} \ \mathrm{years}
\label{tg3}
\eeq
That is, the growth time is short compared to other evolutionary times, e.g. the average accretion spin-up or cooling, but clearly far too long for the  
 instability to  overcome viscosity. Hence, the elliptical instability is also unlikely to be relevant for accreting neutron stars in low-mass x-ray binaries.

The final class of systems which may be of interest are mixed binaries, with a neutron star spiralling into a more massive black hole. In this case, 
the total mass of the system may be much larger, but this does not have much effect on our estimates unless we are dealing with an intermediate mass black hole.   It is easy to see that, cf. \eqref{gt},  the instability growth time increases with the mass ratio. The instability may still act in  mixed binaries but, in order to compensate for the increasing mass ratio, the star will have to spin faster. Of course, we cannot say if this is realistic, as we have not yet observed a pulsar with a black hole companion. 

\section{The p-g instability}

As our final example of dynamical mechanisms that may affect the binary signal, we consider the so-called p-g instability. It has been suggested that high order p- and g-modes of the star couple strongly to the equilibrium tide \cite{wab}. This is a (supposedly) non-resonant mechanism, which means that it would not be restricted to a set frequency range. The available estimates \cite{ess} indicate that the instability sets in at  $f\approx 50$~Hz and that it could have severe impact on the emerging signal. Hence, we need to take the possibility seriously. This involves better understanding the underlying mechanism  and exploring exactly what effect the instability may have. While we are not (currently) in a position to comment on the viability of the p-g instability (as it relies on the details of the nonlinear mode coupling \cite{wab}), we  can nevertheless explore to what extent observations may constrain the theory. 

In the spirit of our estimates for the elliptical instability, we follow \cite{ess} and assume that the p-g instability grows at a rate;
\beq
{1\over t_\mathrm{pg}} \approx 2 \lambda \epsilon \Omega_0 = {2\lambda q \over 1+q} \left( {R_1^3 \over GM_1} \right)^{1/2} \Omega^2 
\eeq
i.e. for our canonical system we have
\beq
{1\over t_\mathrm{pg}}  \approx 7\lambda \left({ f\over 100\ \mathrm{Hz}}\right)^2 \ \mathrm{s}^{-1}
\eeq
where the tidal bulge $\epsilon$ is given by \eqref{bulge}.
We assume that the instability acts above $f\sim 50$~Hz, although the exact cut-off frequency is uncertain (it depends on the damping of high order g-modes, and the star's magnetic field may also have significant influence as these are short lengthscale modes). The function $\lambda(a)$ is slowly varying and expected to lie in the range $0.1-1$ \cite{ess}.

Let us, first of all, establish that the instability grows fast enough to be relevant. As in the case of the elliptical instability, this involves making sure that the growth time is short compared to the inspiral. In this case, we have
\beq
{t_D \over t_\mathrm{pg}} \approx {5\lambda \over 48} {1\over (1+q)^{2/3} } \left( {c^2 R_1 \over GM_1}\right)^{5/2} \left( {\Omega_0 \over \Omega} \right)^{2/3} \approx 42 \lambda \left( {100\ \mathrm{Hz} \over f} \right)^{2/3}
\eeq
That is, for the suggested range of values for $\lambda$, the instability should grow significantly during the inspiral (especially since it does not require a specific resonant frequency match). 

Of course, the instability also needs to overcome viscous damping. In this case, we are dealing with high order modes so we use (it is worth noting that \cite{wab} arrive at essentially the same conclusion through a slightly different scaling argument)
\beq
t_\mathrm{sv} \approx 4\times 10^{-4} \left( {10^3 \over n}\right)^2  \left({M_1\over 1.4\,M_\odot}\right)^{-5/4}
\left({R_1\over 10\mbox{ km}}\right)^{13/4} \left( {T \over 10^6 \ \mathrm{K}}\right)^2 \ \mathrm{s}
\eeq
which leads to the condition 
\beq
{t_\mathrm{sv} \over t_\mathrm{pg}} \approx 3\times 10^{-3} \lambda \left( {10^3 \over n}\right)^2  \left( {f \over 100\ \mathrm{Hz}}\right)^2 \left( {T \over 10^6\ \mathrm{K}}\right)^2 > 1
\eeq
We learn that, even though the p-g instability acts through very high overtone oscillations it may  overcome viscosity throughout much of the inspiral. Specifically, the instability would grow provided
\beq
f > {20 \over \lambda^{1/2}} \left( {n\over 10^3}\right) \left( {T\over 10^6\ \mathrm{K}}\right)^{-1}\ \mathrm{Hz}
\label{con4}
\eeq

\begin{figure}[h]
\begin{center}
\includegraphics[width=16cm,clip]{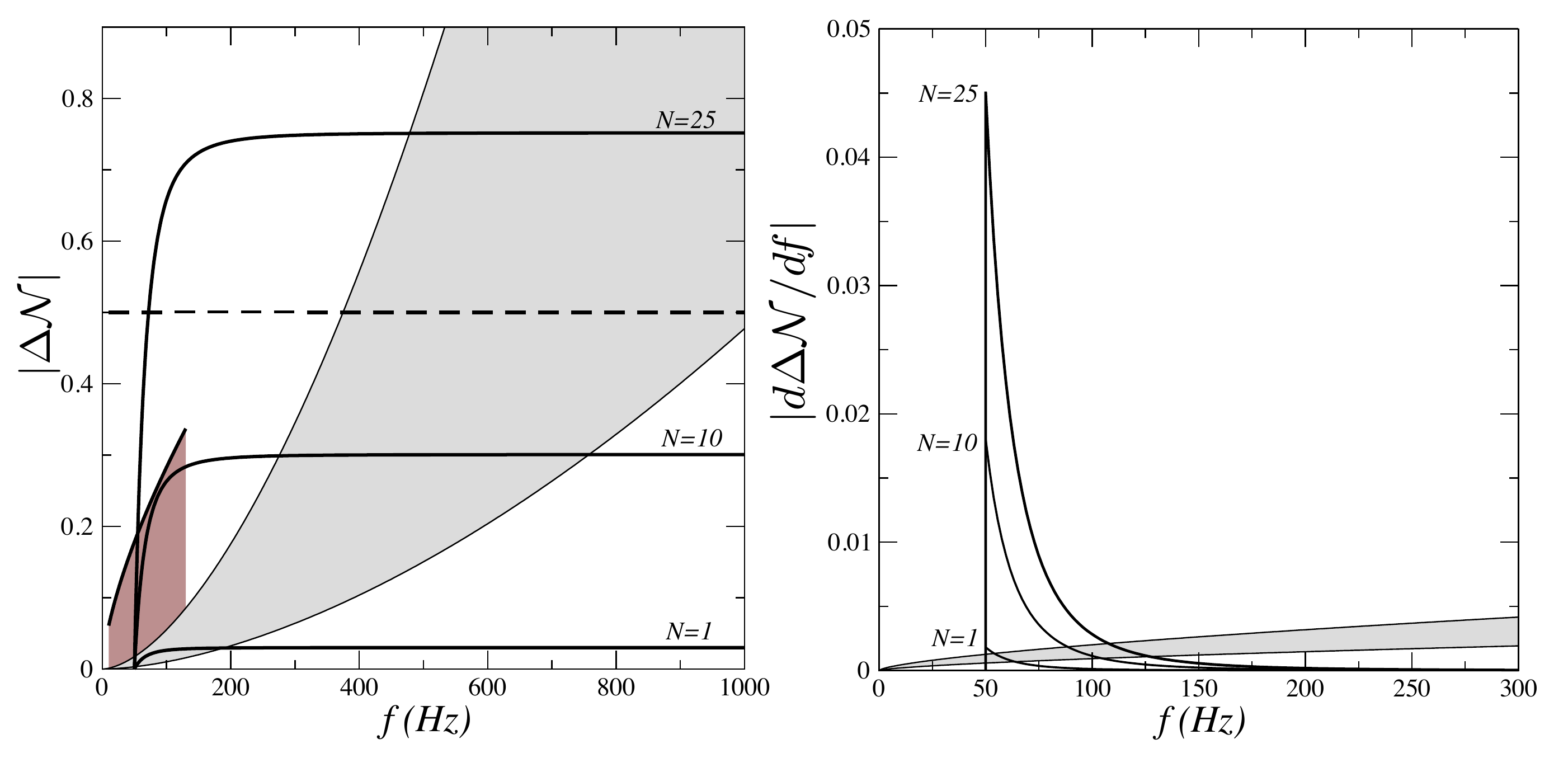}

\end{center}
\caption{A schematic illustration of the impact the p-g instability may have on a binary neutron star signal. Left panel: The estimated shift in the accumulated number of gravitational wave cycles $|\Delta \mathcal N|$ as a function of the gravitational-wave frequency $f$. The grey band recalls the result for the tidal compressibility from Figure~\ref{Love}. The dashed horisontal line represents the indicative level of $|\Delta \mathcal N| \approx 0.5$ above which the effect would leave an imprint in a matched filter search. The solid curves show the estimate effect of the p-g instability ( taking $f=f_b$ in \eqref{dNpg}). The curves represent (taking the unknown parameters $\lambda=\beta=1$ for simplicity), cases where $N=1,10$ and $25$ modes (as indicated) are excited by the instability. Given the uncertainties in the model these estimates should be taken with a fair bit of caution. Finally, the shaded region at low frequencies represents the estimated level of resonant r-modes, given by \eqref{rmode}. Right panel: The same, but for ${|d\Delta \mathcal N/df}|$. This provides a clearer idea of how one can distinguish between the two mechanisms by comparing the effect in a sequence of frequency ranges (time windows). A more realistic model of the p-g instability would account for the finite growth time, which would smooth out the curves, but we would still expect them to be strongly peaked.} 
\label{pg}
\end{figure}

In order to estimate the impact the p-g instability may have on the inspiral signal, we also need to estimate the rate at which energy is drained from the orbit. Again following \cite{ess}, we use
\beq
\dot E_\mathrm{tide} \approx - 10^{-8} { N \beta \over t_\mathrm{pg}} \left( {GM_1^2 \over R_1} \right)
\eeq
where $N$ is the number of unstable modes. The basic idea \cite{ess} is that order $N= 10^3-10^4$ modes may become unstable. This would clearly enhance the impact of the instability, but whether the idea of many modes actually growing to a large amplitude at the same time is viable is not at all clear. The parameter $\beta \le 1$ encodes our ignorance of the saturation mechanism. We need to keep in mind that it could well be that $\beta \ll 1$. 

After a bit of algebra we arrive at
\beq
{\dot E_\mathrm{tide} \over \dot E_\mathrm{gw}} \approx 3\times 10^{-9} {N\beta\lambda \over q(1+q)^{1/3}} \left( {c^2 R_1 \over G M_1} \right)^{5/2} \left( {\Omega_0 \over \pi f} \right)^{4/3} 
\eeq
From this estimate it is clear that one can only expect to constrain the product $N\beta\lambda$ with observations. Combining with $t_D$ and integrating we find that (again, for a canonical equal mass system and assuming that the instability operates from $f=f_0=50$~Hz up to the observed cut-off frequency $f_b$)
\beq
\Delta \mathcal N\approx  - 3\times10^{-2}  N \beta \lambda \left({50~\mathrm{Hz} \over f_0}\right)^3 \left[ 1- \left({ f_0 \over f_b}\right)^3\right]
\label{dNpg}
\eeq
We clearly need the number of excited modes ($N$) to be significant in order for the effect to be detectable, but the conclusion is sensitive to many unknowns. In particular, the estimate for $\Delta \mathcal N$ depends strongly on the frequency at which the instability reaches the saturation amplitude. In essence, while an observed shift in gravitational-wave phasing could be attributed to the p-g instability, it is not clear which part of the (uncertain) theory the absence of the effect would constrain. Another reason to be cautious is illustrated in the right-hand panel of Figure~\ref{pg}. As the most dramatic change in gravitational-wave cycles is associated with a fairly narrow frequency range the effect may be mistaken for a mode resonance.

\section{Concluding remarks}

We have considered three distinct mechanisms involving dynamical tides from an ``observational'' perspective. The motivation for the discussion was to clarify how different aspects of neutron star physics impact on the problem and outline a strategy for how one may be able to constrain this physics with observations. This serves two important purposes. First of all, we gain insight into the extent to which the different mechanisms are within reach of the precise modelling required for a template based gravitational-wave search.  Secondly, we learn that we may be able to make progress even though the details are uncertain. 

As an example, let us consider the case of mode resonances. Our estimates show that such resonances may be difficult to distinguish even with the advanced generation of detectors. However, our understanding is sufficiently uncertain that we cannot rule out the possibility. So let us consider the possible impact on the signal. We know that we are dealing with a resonance, so the impact on the signal should be associated with a specific frequency range. If we were to identify such an imprint, then we would like to identify the specific mode involved (which in turn may constrain the star's internal composition). This could also be tricky, especially if that star is spinning. Given that it may be difficult to obtain tight constraints on the individual rotation rates from the gravitational-wave signal, the rotational corrections to low-frequency modes \eqref{freqshift} introduces significant uncertainty. Nevertheless, one may be able to rule out resonances associated with inertial modes. Given an upper limit on the star's spin, and the fact that an inertial mode must lie in the inertial range \eqref{inert} we know that any associated (quadrupole mode) resonances (associated with the inertial frame frequency) must appear below
\beq
f \approx {2 \Omega_s \over \pi} 
\eeq
In other words, there is a frequency cut-off above which there will be no pure inertial modes. 

Perhaps the most interesting feature one may hope to extract involves modes that are driven unstable by gravitational-wave emission (like the r-modes). In this case one would expect the inspiral to slow down rather than speed up. Such a feature would be tremendously exciting and, as it is a qualitative aspect, may perhaps be within easier reach than a tight constraint on a given resonance. 

Our discussion of the elliptical instability highlighted the possibility of modes forming a nonlinear resonance with the tide. This raises a warning flag, at least in principle, as it introduces complications that have not yet been explored in much detail. The back-of-the-envelope estimates for the elliptical instability suggest that it should be safe to ignore its impact on a binary neutron star signal, but this does not mean that an analogous mechanism could not play a role. We need to keep in mind that the neutron star seismology problem is very rich. 

Finally, we considered the nonlinear and non-resonant p-g instability. Without going into specific detail (which is difficult given the current level of understanding of the mechanism)  we outlined why one may be concerned about this instability acting in an inspiralling binary.  Basically, if the nonlinear coupling between the tide and the
high-order p- and g-modes is, indeed, as strong as been argued \cite{wab}, then the instability would rapidly grow to a large amplitude. If the associated saturation amplitude is sufficiently large, this may leave a detectable imprint on the inspiral signal. However, as we have highlighted, this argument involves a number of unknowns. In order to address these uncertainties the underlying mechanism must be modelled at a more detailed level, but this is difficult given the intricate nature of the nonlinear mode-coupling problem \cite{arras,pantelis} and the need to model the neutron star interior at an adequate level of realism. As long as such results are outstanding, we may still consider to what extent the imprint of the p-g instability would be distinct from that of the other mechanisms we have considered. The key to this issue may be associated with the distinct difference in $|d\Delta \mathcal N/df|$, see the right-hand panel of Figure~\ref{pg}.  The effect of the tidal compressibility increases as $f^{2/3}$ while the estimate for the p-g instability falls of sharply as $f^{-4}$ (after reaching saturation). In essence, one ought to be able to tell the difference  by comparing data for a sequence of frequency ranges (time windows).  The sharp feature of the p-g instability should be distinct from the gradual increase associated with the tidal compressibility.  Of course, as the p-g feature is associated with a finite frequency range one may still confuse it with a mode resonance.

Having outlined the different resonant mechanisms and given the results in dimensionless form, our main conclusion is simple. There is a lot of work to be done on the modelling side. However,  the problem involves major unknowns, like the interior composition, the state of matter, nonlinear dynamics etcetera and these will be difficult to resolve. This motivates us to consider the problem from an "observational" point-of-view. As we enter this new era of neutron-star gravitational-wave astronomy, it may well be that observers will soon lend their struggling theory colleagues a helping hand. 

\section*{Acknowledgements}
We would like to thank Tanja Hinderer for helpful discussions. We acknowledge funding from STFC in the UK through grant number ST/M000931/1.

\end{document}